\documentclass[aps,graphicx,reprint,onecolumn,12pt,tightenlines,longbibliography]{revtex4-1}
\usepackage{amssymb,amsbsy,times,fancyhdr,color}
\usepackage{amsmath}
\usepackage{mathrsfs,amsfonts}
\usepackage{latexsym,afterpage}
\usepackage{graphicx,subfigure,multirow}
\usepackage[colorlinks=true, urlcolor=blue, linkcolor=blue, citecolor=blue]{hyperref}

\usepackage{float}

\usepackage{outlines}

\usepackage[T1]{fontenc}

\usepackage{bm}

\usepackage{amsmath}
\DeclareMathOperator{\Tr}{Tr}

\DeclareMathOperator{\arcsinh}{arcsinh}

\usepackage{soul}
\newcommand{\sas}[2]{{#2}}

\begin{document}

\preprint{APS/123-QED}

\title{Non-Hamiltonian Kelvin wave generation on vortices in Bose-Einstein condensates}

\author{Scott A.~Strong}
 \email{sstrong@mines.edu}
\author{Lincoln D.~Carr}%
\affiliation{%
 \sas{Physics Department, Colorado School of Mines}{Department of Physics, Colorado School of Mines, 1523 Illinois St., Golden, CO 80401, USA}\\
}%




\date{\today}

\begin{abstract}
Ultra-cold quantum turbulence is expected to decay through a cascade of Kelvin waves. These helical excitations couple vorticity to the quantum fluid causing long wavelength phonon fluctuations in a Bose-Einstein condensate. This interaction is hypothesized to be the route to relaxation for turbulent tangles in quantum hydrodynamics. The local induction approximation is the lowest order approximation to the Biot-Savart velocity field induced by a vortex line and, because of its integrability, is thought to prohibit energy transfer by Kelvin waves. Using the Biot-Savart description, we derive a generalization to the local induction approximation which predicts that regions of large curvature can reconfigure themselves as Kelvin wave packets. While this generalization preserves the arclength metric, a quantity conserved under the Eulerian flow of vortex lines, \sas{it does introduce}{it also introduces} a non-Hamiltonian structure on the geometric properties of the vortex line. It is this non-Hamiltonian evolution of curvature and torsion which provides a resolution to the missing Kelvin wave motion. In this work, we derive corrections to the local induction approximation in powers of curvature and state them for utilization in vortex filament methods. Using the Hasimoto transformation, we arrive at a nonlinear integro-differential equation which reduces to a \sas{fully}{modified} nonlinear Schr\"odinger type evolution of the curvature and torsion on the vortex line. We show that this \sas{dynamic}{modification} seeks to disperse localized curvature profiles. At the same time, the non-Hamiltonian break in integrability bolsters the deforming curvature profile and simulations show that this dynamic results in Kelvin wave propagation along the dispersive vortex medium.    
\end{abstract}

\pacs{Valid PACS appear here}
\maketitle



\section{\label{sec:Intro} Introduction}

Quantized vortex lines provide the simplest scaffolding for three-dimensional fluid turbulence. While vortex lines and filaments are fundamental to quantum fluids, they also appear as the geometric primitives in a variety of hydrodynamic settings including atmospheric,  aerodynamic, oceanographic \sas{hydrodynamic}{} phenomenon and astrophysical plasmas.~[\onlinecite{Andersen2014IntroductionEquilibrium}] Through an analogy with the Euler elastica, the twist and writhe of filament structures  appear in models of biological soft-matter and are used to explain the dynamics of DNA supercoiling and self-assembly of bacterial fibers.~[\onlinecite{Mesirov2012MathematicalDynamics},\onlinecite{Scott2006EncyclopediaScience},\onlinecite{Shi1999TheSupercoiling}] Recent research utilizes filaments in less exotic settings where a turbulent un-mixing provides motility to phytoplankton that simultaneously protects them from predation and enhances their seasonal survival.~[\onlinecite{Lindemann2017DynamicsCells},\onlinecite{Durham2013TurbulencePhytoplankton}] In quantum liquids, turbulent tangles of vorticity are known to undergo various changes of state. Transition to the ultra-cold regime is marked by a subsiding Richardson cascade resulting in a randomized tangle with no discernible large-scale structure. In this state, turbulent energy is driven by vortex-vortex interactions where reconnection events trigger a cascade of wave motion along the lines. When these interactions become infrequent, the turbulence begins to relax to a configuration where the mobile vortices spend most of their time in isolation. In this paper, we show that vortex lines in perfect barotropic fluids seek to transport bending along their length in an effort to disperse localized regions of curvature. In mean-field Bose-Einstein condensates, this dynamic predicts the generation of Kelvin waves which are capable of insonifying the Bose gas, providing a pathway to relaxation of ultra-cold quantum turbulence.~[\onlinecite{Leadbeater2001SoundReconnections},  \onlinecite{VanGorder2015TheFluid}] 

Our result derives from the Hasimoto transformation of the Biot-Savart description of the flow induced by a vortex line. We begin with a natural parameterization of the vortex about an arbitrary reference point. In this setting, the Biot-Savart integral (BSI) can be calculated exactly, and asymptotic formulae may be applied to get an accurate description of the velocity field in terms of elementary functions. \sas{}{In the analogous setting of electromagnetism, a steady line current plays the role of vorticity and literally induces a field, in this case magnetic, which has a representation given by the BSI.} \sas{This}{Our analysis yields an induced velocity field generating a vortex} dynamic \sas{}{that} preserves the Hamiltonian structure associated with the Eulerian flow of vortex lines\sas{}{.} \sas{and a}{Additionally, the velocity field a}dmits a Hasimoto transformation \sas{that}{which} describes \sas{}{the evolution of} vorticity through its curvature and torsion. The geometric dynamics are provided by a nonlinear, non-local integro-differential equation of Schr\"odinger type that can be reduced to a local differential equation, which is fully nonlinear in powers of curvature. An analytic analysis assisted by symbolic computational tools  indicates that \sas{all}{} higher order contributions reinforce the changes described by the first correction to the lowest order integrable structure. When simulated, we find that  regions of localized curvature disperse\sas{}{,} causing the production of traveling Kelvin waves. In addition, a gain mechanism emerges to support the dispersion process keeping the helical waves from being absorbed into \sas{the}{an} otherwise straight \sas{background}{vortex line}. 

The local induction approximation (LIA) is the lowest order truncation of BSI representations of the velocity field induced by a vortex line and is used to describe the local flow about vortex points.  The Hasimoto transformation (HT) shows that this is an integrable theory of the curvature and torsion dynamics and, it is argued, incapable of modeling energy transfer through Kelvin waves. It is expected that to solicit Kelvin waves models must undermine or, at least, reallocate the conserved quantities of the flow. There are currently two distinct ways to adapt the LIA to accommodate the study of Kelvin waves. One option is to consider perturbations of the Hasimoto transformation which are known to introduce non-locality and non-integrability into the geometric evolution.~[\onlinecite{Majda2002VorticityFlow}] This technique can study wave motion on wavelengths much smaller than the radius of curvature and much larger than the core thickness. Here the focus is on  kink and hairpin formations in classical vortices capable of self-stretching.~[\onlinecite{Klein1991Self-stretchingLine},\onlinecite{Klein1991Self-stretchingSolutions}] The second method is to consider approximations of the Hamiltonian structure within the LIA and is capable of modeling resonant interactions between Kelvin waves. Currently, there are two derivations of this result which predict cascade features and time scalings arising from the kinetics of Kelvin waves.~[\onlinecite{VanGorder2017MotionApproximation},\onlinecite{Laurie2010InteractionSuperfluids},\onlinecite{Boffetta2009ModelingHelium}]  While the exact source of driven Kelvin wave motion is an open topic~[\onlinecite{White2014VorticesCondensates.},\onlinecite{Kozik2010CommentAndLvov},\onlinecite{Lebedev2010Reply:Turbulence}], there is little doubt that vortex plucking through reconnection generates helical wave motion.~[\onlinecite{Fonda2014DirectReconnection.}] Our results do not seek to model the inception of localized regions of curvature and are instead focused on how the fluid responds to curved abnormalities on the vortex\sas{.}{ and are based on the interplay between the characteristic length scales of curvature, arclength and core size, which are highly constrained in the LIA.} Where the LIA asserts that the curvature, $\kappa$, and torsion, $\tau$, obey a cubic focusing nonlinear Schr\"odinger equation, \sas{}{both Hamiltonian and integrable}, our generalization predicts that the function $\psi = \kappa \mbox{Exp}[\int_{0}^{s}ds' \tau]$ obeys the non-Hamiltonian evolution  $i\psi_{t}+\psi_{ss} + |\psi|^{2} \psi/2 + \left[\tilde{\alpha} \psi\right]_{ss} + f(|\psi|)|\psi|^{2}\psi =0$ where $\tilde{\alpha}$ and $f$ are even functions of curvature with related coefficient structures. \sas{}{An evolution is Hamiltonian if it corresponds to a flow induced by a highly structured vector field. For a completely integrable evolution, one has the ability to utilize linearizations throughout the phase space of solutions associated with a Hamiltonian evolution.  The relationship between the flow and the vector field \sas{which induces}{inducing} it demands that the operator of the Hamiltonian evolution be self-adjoint. Our modified evolution is incapable of supporting such a structure and is, consequently, non-Hamiltonian.} \sas{This}{Moreover, its} fully nonlinear correction allows for energy transfer between helical modes of the vortex line, which significantly alters the evolution of the solitons predicted by the \sas{integrable cubic base}{LIA}.  

The body of this paper is organized into three sections. First, in Sec.~(\ref{sec:BSI}), we derive an exact expression for the BSI representation of the velocity field induced by a vortex line which requires the use of incomplete elliptic integrals. Application of asymptotic formulae makes accessible an expansion of the field in powers of curvature. The coefficients in this expansion depend on the characteristic arclength in ratio with the vortex core size and can be tuned to numerical meshes used in vortex filament methods. Second, in Sec.~(\ref{sec:Hasimoto transform}), we apply the Hasimoto transform to describe the effect of the vector evolution in terms of the curvature and torsion. We show that the corrected dynamic is non-Hamiltonian and allows for helical wave dispersion supported by a gain mechanism. Lastly, we simulate the evolution of soliton, breathing\sas{}{,} and ring dynamics. Specifically, we consider the bright curvature soliton that produces Hasimoto's vortex soliton and find that at various degrees of non-Hamiltonian correction the corrective terms seek to disperse bending in the form of helical wave generation along the vortex line. After this we consider breathing and ring states to find that the intuition given by changes to Hasimoto's soliton carries over and that both cases breakdown into helical wave motions. A notable example is that of a helically perturbed vortex ring which experiences far less dispersive deformation and, in this way, appears stable under simulation. \sas{Perhaps this is an indication that vortex line structures prefer to store bending in helical configurations.}{}      
\section{\label{sec:BSI} The Biot-Savart Integral and Local Induction Models}
The special orthogonal group acts to transform real three-dimensional space by committing rigid rotations about a specific axis which is given by the curl operator.~[\onlinecite{Lax2007LinearApplications}] If we define the instantaneous velocity of a fluid continuum over this spatial domain, then curl represents the axis about which a fluid element rotates. Curves that are parallel to the vorticity vector are called {vortex lines}.~[\onlinecite{Saffman1993VortexDynamics}] A vortex filament is the idealization of a vortex tube whose dynamics are characterized by the behavior of the vortex line. The Biot-Savart integral (BSI) is frequently used to model line and filament dynamics which ignore transverse variations to the vortex core. The BSI provides the unique velocity field such that $\nabla \times \bm{v}=\boldsymbol{\omega}$ defines the vorticity~[\onlinecite{Berselli2002SomeEquation},\onlinecite{Callegari1978MotionVelocity}] and can be thought of as the unique left-inverse of the curl operator.~[\onlinecite{Parsley2012TheThree-sphere},\onlinecite{DeTurck2008Electrodynamics3-space}] In accordance with Helmholtz decomposition of $\mathbb{R}^{3}$, the BSI treats the  velocity field as being sourced by vortical elements convolved with Poisson's formula for the Green's function of a stream reformulation of the hydrodynamic problem.~[\onlinecite{Batchelor2000AnDynamics},\onlinecite{Andersen2014IntroductionEquilibrium}] If the evolution of the vortex is given by the Euler equations, then it is known to be arclength conserving Hamiltonian flow and, in this case, the Biot-Savart volume integral reduces to an integral over the vortex line.~[\onlinecite{Khesin2012SymplecticMembranes}]
\begin{figure}[h]
\includegraphics[width=\textwidth]{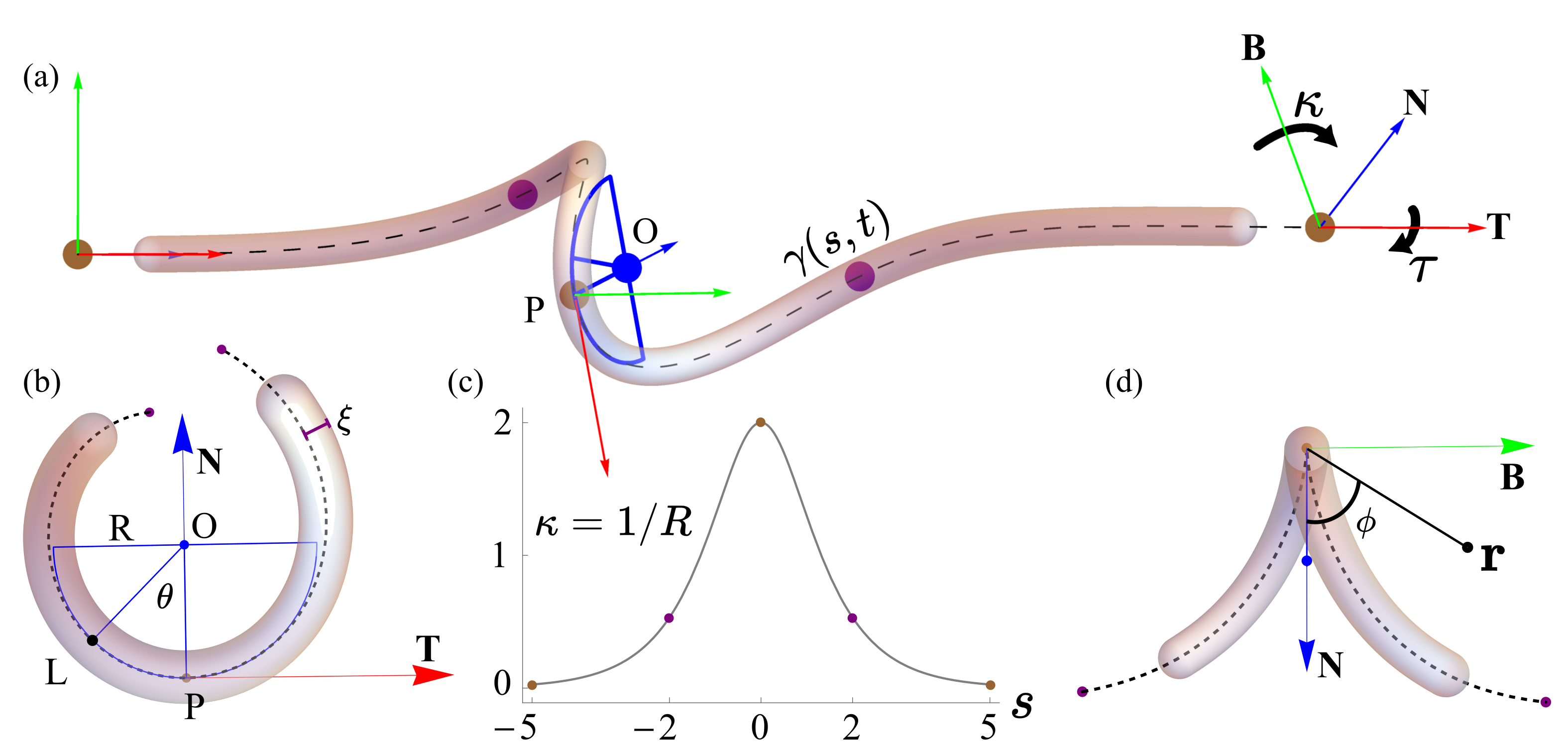}
\caption{Vortex Line Geometry. (a) The local orthogonal frame, tangent (red, $\textbf{T}$), normal (blue, $\textbf{N}$) and binormal (green, $\textbf{B}$) vectors, at the initial, middle and terminal points (brown) of the vortex line, $\bm{\gamma}(s,t)$, with embellished width. At the reference point $P$ we have a local description of the vortex given by a blue semi-circle. Changes to the Frenet frame from point to point are described by the curvature, $\kappa$, and torsion, $\tau$. (b) View\sas{ing}{} down the long axis of $\bm{\gamma}$ where we see the osculating plane spanned by the tangent and normal vectors. The local geometry at $P$ is defined by the radius of curvature $R$, which is related to the curvature by $\kappa=R^{-1}$. The angle $\theta$ sweeps out an arclength from $0$ to $L$ providing a local parameterization to $\bm{\gamma}$ about $P$. The core scale is defined as $\xi$\sas{.}{; in Bose-Einstein condensates taken as the superfluid healing length.} (c) The curvature distribution associated with the vortex line in (a) with unit torsion.  (d) The normal plane is spanned by $\textbf{N}$ and $\textbf{B}$\sas{. An}{, with} observation point, $\textbf{r}=(0,x_{2},x_{3})$, placed in this plane and defined by the polar angle $\phi$.
}
\label{fig:vortex}
\end{figure}
An isolated vortex line is depicted in Fig.~(\ref{fig:vortex}a) and can be defined distributionally for a vortex with homogeneous vorticity density, $\Omega$,
\begin{align}\label{eqn:vortex}
 \bm{\omega}(\textbf{r},t) = \Omega \int_{0}^{L} \delta(\textbf{r}-\bm{\gamma}) \, d\bm{\gamma}, 
\end{align}
where $\bm{\gamma}: (0,L) \times \mathbb{R}^{+} \to \mathbb{R}$, is the dynamic parametric representation of the one-dimensional sub-region on which the vorticity is supported.~[{\onlinecite{Andersen2014IntroductionEquilibrium}}] Additionally, if $\bm{\gamma}=\bm{\gamma}(s,t)$ is parameterized in the natural gauge, then ${\textbf{T}}= \bm{\gamma}_{s}$ is the unit-tangent vector.~[\onlinecite{Burns1991ModernFields}]  Such a distribution of vorticity reduces the BSI to an integral over the vortex line, 
\begin{align}\label{eqn:vortex}
 \bm{v}(\textbf{r},t) = \frac{1}{4\pi} \iiint_{\mathbb{R}^{3}} \frac{\bm{\omega}(\bm{y},t) \times (\textbf{r}-\bm{y})}{|\textbf{r}-\textbf{y}|^{3}}d\bm{y} = - \frac{\Gamma}{4\pi} \int_{0}^{L} \frac{(\textbf{r}-\bm{\gamma})\times d \bm{\gamma}}{|\textbf{r}-\bm{\gamma}^{3}|},
\end{align}
where the circulation/strength $\Gamma$ is the product of $\Omega$ and the characteristic volume resulting from the ideal concentration of vorticity to $\bm{\gamma}$\sas{}{,} and $\textbf{r}$ is the location at which the velocity field is observed in the normal plane, Fig.~(\ref{fig:vortex}d). \sas{}{In the context of electromagnetism, the velocity field plays the role of the magnetic field induced by a steady charge concentrated on the line, $\bm{\gamma}$.} Analogous to problems in electromagnetism, the BSI diverges on the vortex line which is an ideally localized source of the ambient velocity field.  \sas{}{We seek to characterize the flow predcited by Eq.~(\ref{eqn:vortex}). To arrive at a beyond leading order asymptotic representation of the flow predicted by the vortex, multiple layers of analytic work will be needed. To assist the reader we provide an overview of the process.} 

\begin{figure}
\includegraphics[width=\textwidth]{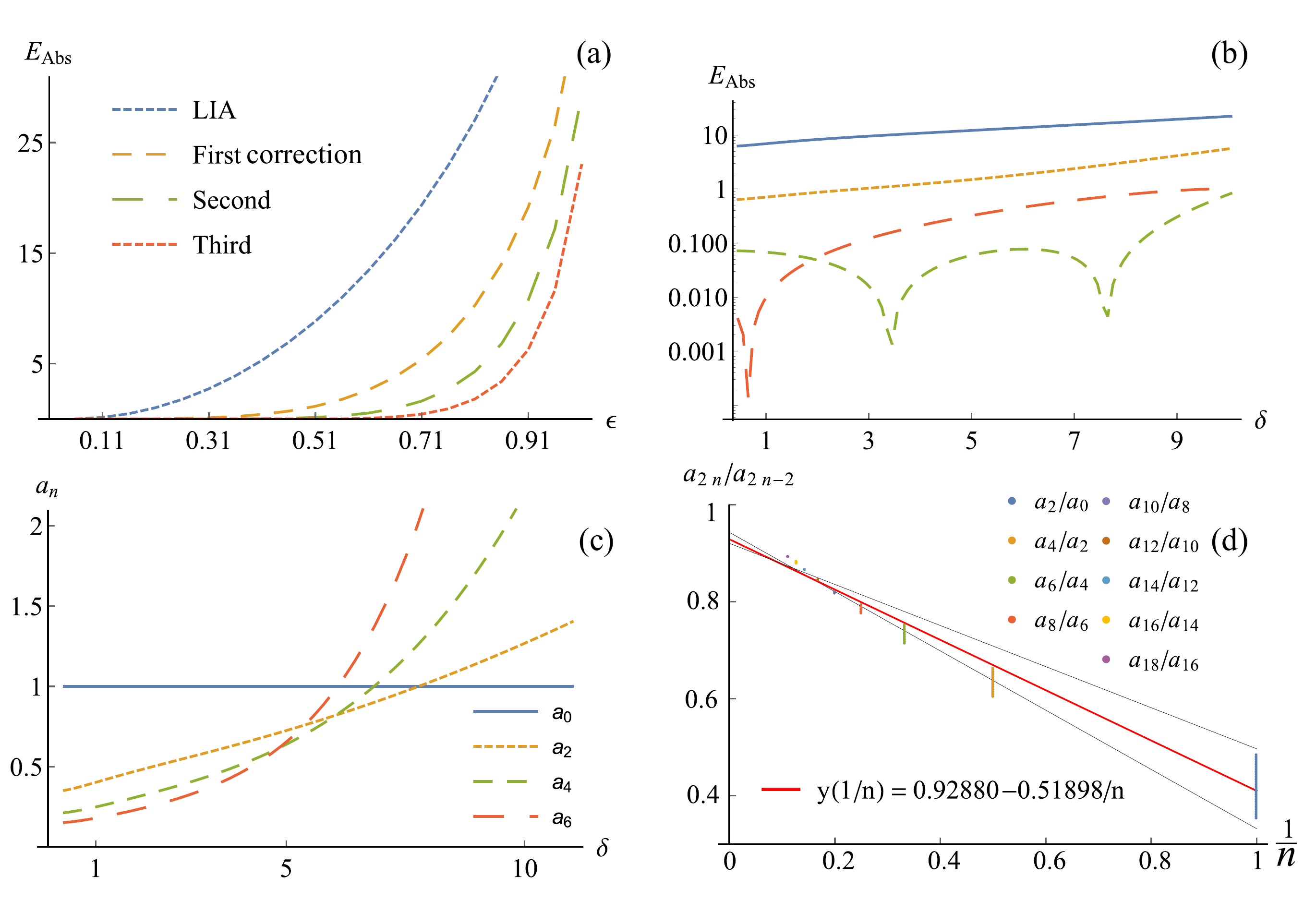}
\caption{Absolute errors and convergence analysis. (a) The average absolute error, $E_{\mbox{Abs}}$, of LIA and the first, second, and third corrections given by expanding the asymptotic representation of the Biot-Savart integral in powers of curvature. The error is calculated against   numerical quadrature converged to six digits of accuracy. While the accuracy improves with higher order corrections, what is noteworthy is how quickly LIA loses  its accuracy for large curvature. (b) The error as a function of $\delta$ shows that LIA is a generally inaccurate approximation.  (c) \sas{Plots t}{T}he first four non-trivial coefficients in \sas{the}{our} expansion \sas{}{correcting LIA} as a function of $\delta$\sas{ which are positive and increasing}. (d)  A Domb-Sykes plot where the data are given by a uniform sampling of the coefficients over $a + 0.01< \delta < 2$\sas{.}{ where $a$ is the Bustamante-Nazarenko cutoff.}  We see that $a_{2}/a_{1}$ varies the most over this sampling. The data for each $\delta$ in the sampling are fitted to linear models with an average square residual of $0.98$. The red curve depicts a line whose vertical intercept and slope are given by averaging over the system of linear fits. The resulting vertical intercept predicts a radius of convergence in $\epsilon$ of roughly $0.94$.}
\label{fig:error}
\end{figure}

\sas{}{The analysis which we apply to Eq.~(\ref{eqn:vortex}) is as follows.} 
After stating a parameterization for $\bm{\gamma}$ one must consider field points asymptotically close to the vortex line with $|\textbf{r}|$ on the order of vortex core size defined by $\xi$, see Fig.~(\ref{fig:vortex}b). \sas{}{Since the field diverges at the vortex source,  we must regularize the BSI which is tantamount to eliminating high frequency oscillations along the vortex.} The local induction approximation (LIA) is the reduction of BSI to its logarithmic singularity. A classic treatment can be found in Batchelor~[\onlinecite{Batchelor2000AnDynamics}] who derives the result by formally setting the ratio of the observation point \sas{}{magnitude}, $|\textbf{r}|$, with the local radius of curvature, $R$, to zero. While this is the most expedient route to the lowest order kinematics, it quickly loses accuracy at moments where the local curvature becomes large, see Fig.~(\ref{fig:error}a). \sas{}{While our exact elliptic representation of the regularized BSI can certainly resolve the field at moments of large curvature, they obstruct our understanding of primitive  wave motions along the vortex understood through Hasimoto's transform. Thus, we apply asymptotic approximations to the exact field to get simpler representations in powers of curvature.} \sas{Such events}{Moments of large curvature} are a \sas{natural}{} consequence of vortex-vortex interactions leading to tent/cusp like formations and, when pushed far enough, reconnection. In LIA the ratio of arclength to vortex core radius is required to be large, $L\gg |\textbf{r}|\sim \xi$, which is incompatible with reconnection where the vorticity local to the event drives the dynamics. \sas{}{We now correct LIA so that we can describe the dynamics in this regime of interest.}

The dynamics nearing reconnection remain unresolved by LIA, and we seek to rectify this issue by retaining curvature dependence in the BSI integrand. While our approximation recovers LIA in the standard limit, it also allows for an interplay between the characteristic length scales of curvature, arclength and core size that is forbidden by LIA and allows for accuracy in scenarios applicable to modeling wave motions post-reconnection. \sas{}{The end result will be that the speed of the local velocity field is given by $|\bm{v}| \propto \alpha(\kappa)$, where}
\begin{align}\label{eqn:alpha}
 \alpha(\kappa) = \sum_{n=0}^{\infty} a_{2n}\kappa^{2n},
\end{align}
\sas{}{such that restricting the series to $a_{0}$ results in the LIA. Prior to regularization and asymptotic approximation we extract the divergent component of BSI.} First, we restrict the parameterization of the vortex line at an arbitrary point, $P$, in the osculating plane, see  Fig.~(\ref{fig:vortex}b), and consider only the binormal component of the local velocity field to get 
\begin{align}\label{eqn:binormal1}
 v_{3}(\textbf{r})  \propto \int_{0}^{L} \frac{\epsilon_{3jk}(x_{j}-\gamma_{j})d\bm{\gamma}_{k}}{|\textbf{r}-\bm{\gamma}|^{3}} = -x_{2}\int_{0}^{L} \frac{\gamma_{1}'}{|\textbf{r}-\bm{\gamma}|^{3}}\, ds   +\int_{0}^{L} \frac{\gamma_{2}\gamma_{1}'-\gamma_{1}\gamma_{2}'}{|\textbf{r}-\bm{\gamma}|^{3}}\, ds,
\end{align}
where we have made use of the Levi-Civita permutation tensor\sas{}{, $\epsilon_{ijk}$,} in conjunction with the Einstein summation convention to compute the cross-product integrand in terms of the components of the parameterization, $\bm{\gamma} = (\gamma_{1}, \gamma_{2},0)$ and their first partials with respect to arclength, $\partial_{s}\bm{\gamma} = (\gamma_{1}', \gamma_{2}',0)$.  Additionally, we may omit the first\sas{-}{}component in the location of the observer, $\textbf{r}=(0,x_2,x_3)$, who is restricted to the normal plane, see Fig.~(\ref{fig:vortex}d). A quick computation of ${v}_{2}$ reveals the circulatory counterpart to the $x_{2}$ term of the velocity field. Thus, the last term in Eq.~(\ref{eqn:binormal1}) defines a non-circulatory binormal flow, which is understood as a non-stretching dynamic capable of producing  geometric alterations to the vortex line.

To derive a locally induced flow from Eq.~(\ref{eqn:binormal1}) one must consider field points asymptotically close to the vortex line and also regularize the divergence by excising a portion of the vortex line neighboring the reference point. The length of the excised domain of integration in Eq.~(\ref{eqn:binormal1}) is often decided in an ad hoc manner.~[\onlinecite{Schwarz1985Three-dimensionalInteractions}]  However, the recent work of Bustamante and Nazarenko provides a regularization cutoff consistent with the mean field vortices of a Bose-Einstein condensate.~[\onlinecite{Bustamante2015DerivationEquation}] To make use of this we specify a parameterization of the vortex line local to the reference point and explicitly  process the binormal flow in Eq.~(\ref{eqn:binormal1}).  In generalization to Batchelor's work~[\onlinecite{Batchelor2000AnDynamics}], we consider a dynamic element of vorticity given by $\bm{\gamma}(s,t) = (R \sin(\kappa s), R - R \cos(\kappa s),0)$ where $\kappa^{-1} = R=R(s,t)$ and $\kappa s= \theta \in (-\pi,\pi)$, see Fig.~(\ref{fig:vortex}b), whose quadratic approximation is consistent with the parameterization given in ~[\onlinecite{Batchelor2000AnDynamics}]. In either case, the  exact representation of the induced field is given in terms of incomplete elliptic integrals. Since our derivation relies on differentiation of the integral with respect to an internal parameter, elliptic integrals of both first and second kind appear.~[\onlinecite{Strong2012GeneralizedTurbulence}] Integrating the final term in Eq.~(\ref{eqn:binormal1}) over the angle $\theta$\sas{}{,} which is related to arclength by $s=R\theta$, gives the following representation for the binormal flow induced by a plane circular arc, 
\begin{align}\label{eqn:binormal2}
\bm{v}_{\textbf{B}} = -\frac{\Gamma \kappa}{4\pi} \int_{0}^{L} \frac{\cos(\theta) - 1}{(c_{1} + c_{2} \cos(\theta))^{3/2}} \, d\theta, \quad c_{1} = \epsilon^{2}  -2 \epsilon \cos(\phi)+2, \quad c_{2} = 2\epsilon \cos(\phi)-2,
\end{align}
%
where $\phi$ is the polar angle of the field point in the normal plane, see Fig.~(\ref{fig:vortex}d), and $\epsilon = |\bm{r}| \kappa$ is the product of the distance of the field point and the local curvature. Except at moments of reconnection where cusps form on the vortex line, this parameter is small, though not formally zero as in Batchelor's derivation. As was perhaps first witnessed with the theory of boundary layers, the predictions in the asymptotic regime of $\epsilon \to 0$ differ significantly from those stemming from $\epsilon=0$, which prohibits the existence of large curvatures. The corresponding indefinite integral can be evaluated to the form

\begin{align}\label{eqn:IntegralEllipticCircle}
\int \frac{\cos(\theta) - 1}{(c_{1} + c_{2} \cos(\theta))^{3/2}} \, d\theta = F  \, C_{-}-E  \, C_{+}+\frac{c_{2} \sin (\theta)}{c_{2} (c_{1}-c_{2}) \sqrt{c_{1}+c_{2} \cos (\theta )}}
\end{align}
where
\begin{align}\label{eqn:coefficients}
&F=F\left(\left.\frac{\theta }{2}\right|\frac{2 c_{2}}{c_{1}+c_{2}}\right), \\
&E=E\left(\left.\frac{\theta }{2}\right|\frac{2 c_{2}}{c_{1}+c_{2}}\right), \\
&C_{\mp}=\frac{2(c_{1}\mp c_{2}) \sqrt{\displaystyle\frac{c_{1}+c_{2} \cos (\theta )}{c_{1}+c_{2}}}}{c_{2} (c_{1}-c_{2}) \sqrt{c_{1}+c_{2} \cos (\theta )}}.
\end{align}
\sas{where}{Here} we have used the traditional notation for the elliptic integrals native to Mathematica, which are related to the standard notations by $F(z,m) = F(z|m^{2})$ and $E(z,m) = E(z|m^{2})=\int_{0}^{z} \sqrt{1-m^{2} \sin^{2}(t)}dt$.  Of the terms in Eq.~(\ref{eqn:IntegralEllipticCircle}), only the first is divergent as $\epsilon \to 0$.  There are various asymptotic formulae for elliptic integrals.~[\onlinecite{ToshioFukushima2012SeriesIntegrals}] The result of Karp and Sitnik was found to be more accurate than the prior result of Carlson and Gustafson, in the sense of average absolute and relative errors, over a wider range of parameters.~[\onlinecite{Karp2007AsymptoticSingularity},\onlinecite{Carlson1994}] As $\epsilon\to 0$, the Karp and Sitnik representation of the divergent term in Eq.~(\ref{eqn:IntegralEllipticCircle}) is given by
\begin{align}\label{eqn:KS}
 F\left(\left.\frac{\theta }{2}\right|\frac{2 c_{2}}{c_{1}+c_{2}}\right)C_{-}\sim \mathcal{A} \left[ \ln\left(\frac{\epsilon  \sin\left(\theta/2\right)}{\sin(\theta/2)\sqrt{A_{3}}+\sqrt{A_{2}}}\right) -\frac{2A_{3}}{\sqrt{A_{2}}\sin(\theta/2)} \ln\left(\frac{\sqrt{4A_{1}}}{\sqrt{A_{1}}+2}\right)\right],
\end{align}	
where
\begin{align}\label{eqn:KSAux}
A_{1} &= \cos(\theta) +1 ,\\
A_{2} &= c_{2} \cos(\theta) - c_{1},\\
A_{3} &= 2c_{1} - \epsilon^{2},\\
D_{1} &= A_{3} (A_{3}-c_{1})/2,\\ 
\mathcal{A} &= \frac{A_{2}}{2D_{1}A_{3}^{3/2}}.
\end{align}
Comparing this approximation to the elliptic form gives an average absolute and relative error of less than 1.5\% over the parameter domain $(\epsilon, \theta) \in (0,1) \times (10^{-9},\pi)$, which gets significantly better away from the boundaries in $\theta$ and away from the upper bound in $\epsilon$. Noting that \sas{as $\epsilon{\to} 0$}{for $\epsilon{\ll} 0$} , $c_{1} \sim 2$, $c_{2} \sim -2$, $A_{2}\sim -2A_{1}$ and $A_{3} \sim 4$, implies the divergence manifests from the $\ln(\epsilon)$ term in Eq.~(\ref{eqn:KS}). However, discarding the remaining terms in Eq.~(\ref{eqn:KS}) raises both error measures to roughly 30\%. Thus, our interest is in how terms other than $\ln(\epsilon)$ temper the divergence away from the singularity\sas{.}{and how we can incorporate their effects into our expansion of the field strength as a function of curvature given by Eq.~(\ref{eqn:alpha}). }However, before we proceed, we must regularize the integral by omitting a portion of vortex neighboring the reference point.     
%
%

Previously, regularization of the BSI for a vortex line were conducted ad hoc with cutoffs tuned to experimental observation.~[\onlinecite{Moore1972TheFlow}] However, a recent derivation of BSI from the Gross-Pitaevskii (GP) equation, which models mean--field Bose--Einstein condensates, provides an a priori regularization of high-frequency spatial modes.~[\onlinecite{Bustamante2015DerivationEquation}] Adapting the results to our parameterization defines a domain of integration, ${D} = (a \epsilon, \epsilon \delta)$, such that $\pi > \epsilon \delta = \kappa L$ where $a\approx 0.3416293$ and $L$ is the length of half  of the symmetric vortex arc. \sas{}{It is important to note that the cutoff parameter, $a$, of Bustamante and Nazarenko is not phenomenological as in prior regularization techniques. Instead, by reformulating the Hamiltonian structure of the Gross-Pitaevskii equation to be consistent with vortex lines, the authors were able to numerically approximate a cutoff value under the assumption that density fluctuations ceased far from the vortex. Interestingly, their derivation of a self-consistent cutoff relies on incorporating the leading order contributions of quantum pressure and the mean-field potential, in addition to the kinetic term of the Bose-Einstein condensate Hamiltonian. Thus, our generated Kelvin waves arise from a regularization procedure that takes into account a non-trivial portion of vortex core dynamics.} 

Assisted by Mathematica, we compute the coefficients of a power-series expansion of Eq.~(\ref{eqn:KS}) up through the first twenty terms. Due to the complicated dependence on $\epsilon$, computation of higher order coefficients requires more sophisticated handling of system memory. Restricting our field observation to the normal plane and averaging over $\phi$, we find that the first ten odd coefficients vanish, which is consistent with the symmetry properties of known matched asymptotic expansion.~[\onlinecite{Svidzinsky2000DynamicsCondensate}] \sas{For t}{T}he \sas{}{two-}parameter regime, \sas{$(\epsilon,\delta)\in(0,0.35)\times(a,10)$}{$0<\epsilon<0.35$ and $a < \delta < 10$}, \sas{}{corresponds to vortex lines whose radius of curvature is up to 35\% of the core radius that is integrated from the Bustamante-Nazerenko cutoff through an order of magnitude greater than the core radius.} the following regularized two-term approximation has an average absolute error that is roughly $3.1\%$ different than the exact elliptic form, 
\begin{align}
\label{eqn:LIA2}\alpha^{*} &= \int_{a\epsilon}^{\delta \epsilon} \frac{\cos(\theta) -1 }{(c_1 + c_2 \cos(\theta))^{3/2}}d\theta \sim a_{0} + a_{2} \epsilon^{2},\\
a_{0}(\delta,a)& = g(a)-g(\delta),\\
 a_{2}(\delta,a)& = \frac{h(a)}{48}-\frac{h(\delta)}{48}+ \left(\frac{3}{4}\right)^{2}a_{0},  
\end{align}
where $g(\eta)= \arcsinh(\eta)$ and $ h(\eta) =(\eta^5-12\eta^3-11\eta) (1+\eta^{2})^{-3/2} $. \sas{}{We define $\alpha$, from Eq.~(\ref{eqn:alpha}), by normalizing all coefficients of Eq.~(\ref{eqn:LIA2}) by $a_{0}$. That is, $ \alpha=\alpha^{*}/a_{0}$.}

\sas{}{Prior to taking Hasimoto's transformation, we would like to make sure that our expansion recovers LIA. Also, we would like to understand how well the beyond lowest order LIA terms in Eq.~(\ref{eqn:alpha}) approximate the BSI defined by Eq.~(\ref{eqn:binormal2}).} Neglecting the quadratic term and noting that $\arcsinh(\delta) \sim \ln(\delta)$ \sas{as $\delta{\to} \infty$}{for $\delta{\gg} 1$} and  yields $\alpha^{*} \sim -\ln(\delta) + O(1)$\sas{,}{} \sas{i.e.~}{That is, from Eq.~(\ref{eqn:LIA2}) we  recover} the standard LIA with $\delta = L/|\textbf{r}|$. In Fig.~(\ref{fig:error}a) we depict the absolute percent error, averaged over $0.1<\delta<10$ for the first, second and third corrections as a function of $\epsilon$. \sas{}{Specifically, $\alpha^{*} \approx a_{0}$ gives the LIA, while the higher order corrections correspond to  $\alpha^{*} \approx a_{0} + a_{2} \epsilon^{2}$ (First correction), $\alpha^{*} \approx a_{0} + a_{2} \epsilon^{2}+a_{4} \epsilon^{4}$ (Second), and $\alpha^{*} \approx a_{0} + a_{2} \epsilon^{2}+a_{4} \epsilon^{4}+a_{6} \epsilon^{6}$ (Third).}  We see that the terms in Eq.~(\ref{eqn:KS}) tempering the logarithmic singularity significantly improves the lowest order approximation   for regions with larger curvature. Figure (\ref{fig:error}b) provides a logarithmic plot of this error now as a function of $\delta$ and averaged over $0.05<\epsilon<1$. We see that LIA is  an inaccurate approximation of the binormal speed of the vortex when curvature is large. In Figs.~(\ref{fig:error}c) and (\ref{fig:error}d) we plot information about the coefficients in our expansion of the asymptotic representation of the local velocity field. The Domb-Sykes plot~[\onlinecite{Hinch1991PerturbationMethods},\onlinecite{Georgescu1995AsymptoticEquations}] in Fig.~(\ref{fig:error}d) shows that we expect a mean radius of convergence for the series of approximately $\epsilon=0.94$, which is consistent with the assumptions of our approximation. Altogether, we find that the LIA demands the lowest order term dominates the representation and requires the scale separation,  $\epsilon \ll \epsilon\delta \ll \delta=L/|\bm{r}|$\sas{,}{In contrast,} our generalized local induction equation permits the study of flows where the scale separation is much less restrictive. \sas{Specifically, it requires {the} vortex arcs {have} a small enough central angle, $\theta$, such that $\epsilon < \epsilon \delta < \pi$ where $\epsilon < 1$.}{Specifically, it requires that the vortex arc has a small central angle, $\theta$, and a radius of curvature bounded by vortex core size,}
\begin{align}
 \epsilon \delta &< \pi, \\ \epsilon &< 1.
\end{align}
For a barotropic inviscid fluid, Kelvin's circulation theorem tells us that a vortex line flows as if frozen into the ambient fluid flow. Thus, the contortions it undergoes result from the flow which it induces. Furthermore, if the fluid is incompressible and of infinite extent, then its autonomous dynamics are completely determined by the Biot-Savart integral, which provides a representation of the ambient flow. Since the vortex line inherits the velocity of the fluid background, we have the following evolution law for the vortex, 
\begin{align}\label{eqn:BNF}
\frac{\partial \bm{\gamma}}{\partial t} = \frac{\Gamma a_{0} }{4\pi} \alpha(\kappa) \frac{\partial \bm{\gamma}}{\partial s}\times \frac{\partial^{2} \bm{\gamma}}{\partial s^{2}},
\end{align}
where $\alpha = \alpha^{*}/a_{0}$, given by Eq.~(\ref{eqn:LIA2}), is even in the curvature variable and $\bm{\gamma}_{s}\times\bm{\gamma}_{ss}={\textbf{B}}$. The local induction approximation is then the linear approximation to $\alpha$, in $\kappa$, where $a_{0} = \ln(\delta)$ and the starting point of Hasimoto's mapping. In the next section, we apply this transformation to a generalization of Eq.~(\ref{eqn:BNF}) and show that higher order curvature effects break the fragile integrability and allow the vortex medium to support a wider array of nonlinear waves. 

\section{\label{sec:Hasimoto transform} Hasimoto's Transformation of Binormal Flows}
A space curve is defined by how the local tangent, normal and binormal frame, $(\textbf{T},\textbf{N},\textbf{B})$, changes between points, see Fig.~(\ref{fig:vortex}). The Frenet-Serret equations, in the natural arclength parameterization, is a system of first-order skew-symmetric ordinary differential equations that recovers the local frame based on how a curve fails to be straight (curvature, $\kappa$) and planar (torsion, $\tau$) along its arclength.~[\onlinecite{Kuhnel2006DifferentialManifolds}]
How Eq.~(\ref{eqn:BNF}) affects the global geometry is described by the Hasimoto transformation.~[\onlinecite{Hasimoto1972}] The Hasimoto transform is a coordinate change which decouples the evolution of the extrinsic shape defined by the parametric curve from its intrinsic curvature and torsion, with the Frenet-Serret equations acting as the intermediary. The modern perspective is that the Hasimoto transform is a scalar manifestation of the Madelung transformation, which is the inversion of volume preserving mappings from the Euler equation phase space to the projective space of non-vanishing complex functions.~[\onlinecite{Khesin2017GeometricTransform}] In light of Bustamante and Nazarenko's work, the geometric analysis provided by Hasimoto transform applied to BSI flows is formally a study of isolated vortex lines in Bose-Einstein condensates. This analysis also describes Eulerian fluids whose phase space is made topologically distinct from the isotropic state with trivial vorticity through the presence of vortex lines.

The Euler evolution of vortex lines is known to be a Hamiltonian flow of the arclength metric. Shortly after Hasimoto's discovery, it was recognized that the velocity fields defined by LIA were Killing, or arclength preserving, on $\mathbb{R}^{3}$ and that Hasimoto transform connects  them to the sequence of commuting Hamiltonian flows of the integrable cubic focusing nonlinear Schr\"odinger equation.~[\onlinecite{Langer1991PoissonEquation},\onlinecite{Langer1990TheCurves}]  The Hasimoto evolution complicates itself substantially when perturbing off the Killing structure, leading to a quasilinear integro-differential equation of Schr\"odinger type, it also produces other mixtures \sas{}{of} Schr\"odinger and Korteweg-de Vries hierarchies.~[\onlinecite{Arnold1999TopologicalHydrodynamics}, \onlinecite{Majda2002VorticityFlow},\onlinecite{Fukumoto1991Three-dimensionalVelocity}] Generalizing LIA in powers of curvature maintains the Killing structure and, consequently, the Hamiltonian of the Euler equations. The cost, however, is that it introduces a non-Hamiltonian evolution to the  geometric variables of curvature and torsion. In other words, nonlinear curvature dependent binormal flow is an arclength preserving non-Hamiltonian flow on the vortex geometry that gives rise  to the \sas{non-Hamiltonian}{dispersive} bending \sas{}{generation} along the vortex.  

If the Madelung transformation~[\onlinecite{Madelung1926EineSchrodinger}] describes a mean-field \sas{GP}{Gross-Pitaevskii} \sas{BEC}{Bose-Einstein condensate} as a perfect and incompressible fluid in which rotation must manifest through circulation about a topological defect known as a vortex line, then the Biot-Savart integral provides its evolution, in the absence of boundary effects. We consider the Hasimoto transformation on a perturbation of our Biot-Savart derived binormal flow, Eq.~(\ref{eqn:BNF}),
\begin{align}\label{eqn:Hasimoto transform1}
\frac{\partial \bm{\gamma}}{\partial t} =  A \alpha(\kappa) \kappa \textbf{B}+\mu \nu
\end{align}
where $A=\Gamma a_{0} / 4\pi$, $\bm{\nu}\in\mathbb{R}^{3}$ and $\mu \ll 1$. Perturbations of this form were first considered by Klein and Majda and will be a useful contrast to our result.~[\onlinecite{Majda2002VorticityFlow}] Specifically, we see that arclength metric preserving modifications of LIA generally result in gain/loss mechanisms. First, however, we express two key quantities in Hasimoto's work \sas{}{that track the frame changes from point to point and through time,}
\begin{align}
\label{eqn:HasNormal} \bm{\aleph}(s,t) &= ({\textbf{N}}(s,t) + i {\textbf{B}}(s,t))e^{i\phi(s,t)},\\
\label{eqn:HasWave} \psi(s,t) &= \kappa(s,t)\,  e^{i \phi(s,t)},
\end{align}
\sas{}{which are written in} in terms of the dimensionless phase, $\phi(s,t) = \int_{0}^{s} ds' \tau(s',t)$. The Hasimoto frame, $({\textbf{T}},\bm{\aleph}, \bar{\bm{\aleph}})$, constitutes an orthogonal coordinate system for $\mathbb{R}\times\mathbb{C}^{2}$ with respect to the Hermitian inner-product. Our derivation is greatly simplified by using the following modifications of the standard commutator and anti-commutator operators, $\left[A,B\right]= A\bar{B}-\bar{A}B$ and $\left\{A,B\right\}= A\bar{B}+B\bar{A}$, which takes into account the way complex conjugation \sas{}{, $\bar{\bm{\aleph}}=({\textbf{N}} - i {\textbf{B}})e^{-i\phi},$} appears in our adaptation of Hasimoto's transformation. \sas{}{Additionally, we will use subscript notation to denote partial differentiation, $\partial_{s} \psi = \psi_{s}$.}
\sas{The first notable change to the process, due to Eq.~(), occurs when considering the dynamics of the local tangent vector,}{The first notable change to Hasimoto's process occurs when trying to express the derivative of Eq.~(\ref{eqn:Hasimoto transform1}) in terms of $\bm{\aleph}$. In Sec.~(\ref{sec:simulations}) we introduce the Frenet-Serret equations, Eq.~(\ref{eqn:FS}). For now, we note that $\textbf{B}_{s}=-\tau \textbf{N}$ which, in conjunction with Eq.~(\ref{eqn:Hasimoto transform1}) and the relation  $\left(\bm{\gamma}_{t}\right)_{s}=\textbf{T}_{t}$, gives}
\begin{align}\label{eqn:Hasimoto transform2}
{\textbf{T}}_{t} = \frac{i A}{2}\left[\eta, \bm{\aleph}\right] + \mu \bm{\nu}_{s}, \quad \eta = \frac{\partial}{\partial s}\left(\alpha \psi\right)=\left(\alpha \psi\right)_{s},
\end{align}
\sas{}{where we have assumed continuity of the second order mixed space-time partial derivatives so that $\partial_{st}=\partial_{ts}$.}
 This alteration affects a critical step in the transformation where two definitions of the mixed partial derivative of $\bm{\aleph}$ are equated. Specifically, we have the two equations
\begin{align}
\label{eqn:Hasimoto transform3} \bm{\aleph}_{st}&=-\psi_{t}{\textbf{T}} - \psi {\textbf{T}}_{t},\\
\label{eqn:Hasimoto transform4} \bm{\aleph}_{ts}&=iR_{s} \bm{\aleph}-i\left( R \psi +   A \eta_{s}\right) {\textbf{T}} - \frac{iA\eta}{2} \left\{ \psi,\bm{\aleph} \right\},
\end{align}
where the first expression derives from the definition of the frame coupled to Eq.~(\ref{eqn:Hasimoto transform2}) and the second from the orthogonal decomposition of ${\bm{\aleph}_{t}}$. Projecting out the coefficients using the local tangent gives,
\begin{align}
\label{eqn:Hasimoto transform5} i\psi_{t} + A(\alpha \psi)_{ss} + R \psi + i \psi\mu (\bm{\nu}\cdot {\textbf{T}}) =0,
\end{align}
while using the Hasimoto normal vector gives
\begin{align}
\label{eqn:Hasimoto transform6} R_{s} = \frac{A}{2}\left\{ \psi,\eta \right\}  + \frac{i \mu}{2} \left[\psi, \bm{\aleph}\right]\cdot \bm{\nu}_{s}.
\end{align}
Letting $\alpha=1$ implies that $\eta=\psi_{s}$ and if $\mu=0$, we have Hasimoto's original transformation where the first-term becomes the exact derivative of $A|\psi|^{2}/2$. However, after integrating by parts to find $R$ we have that, up to constants of integration, Eq.~(\ref{eqn:Hasimoto transform5}) is generally given by\sas{,}{the integro-differential equation,}
\begin{align}
\label{eqn:Hasimoto transform7} i\psi_{t} + A(\alpha \psi)_{ss} +  \frac{A\psi}{2} \int_{0}^{s} \alpha_{s'}|\psi|^{2} ds'  + \mu \left\{ i[(\bm{\aleph}\cdot \bm{\nu}_{s})_{s}-\psi\nu_{s}\cdot {\textbf{T}}] +\psi \int_{0}^{s} \mbox{Im}[\psi \bar{\bm{\aleph}}] \cdot \bm{\nu}_{s'}\,ds'\right\}=0.
\end{align}
\sas{which is a fully nonlinear integro-differential equation of Schr\"odinger type.}{As this evolution contains nonlinearities in the highest order derivative, it is fully nonlinear. Additionally, we say that it is of Schr\"odinger type since $\alpha\to 1$ and $\mu \to 0$ produces the cubic focusing nonlinear Schr\"odinger equation consistent with LIA.} The terms associated with $\mu$ were found to model perturbations whose  wavelength was small with respect to the radius of curvature, but long compared to core thickness. Letting $\alpha=1$ and $\bm{\nu}=\mu \textbf{B}$ reduces Eq.~(\ref{eqn:Hasimoto transform7}) to a complex Ginzburg-Landau type equation with a torsion driven gain/loss term and shows that even the simplest arclength preserving alteration to LIA is capable of breaking its fragile integrability. Focusing now on ambient flows completely characterized by Eq.~(\ref{eqn:BNF}), we let $\mu=0$ and expand $\alpha$ in powers of curvature to calculate the first integral in Eq.~(\ref{eqn:Hasimoto transform7}) explicitly. Doing so, under an appropriate time rescaling, reduces Eq.~(\ref{eqn:Hasimoto transform7}) to the fully nonlinear differential equation,
\begin{align}\label{eqn:Hasimoto transform8}
i\psi_{t}+\psi_{ss} + \frac{1}{2} |\psi|^{2} \psi + \left(\tilde{\alpha} \psi\right)_{ss} + f(|\psi|)|\psi|^{2}\psi =0,
\end{align}
such that $\alpha = 1 + \tilde{\alpha}$ and 
\begin{align}\label{eqn:Hasimoto transform8aux1}
f(|\psi|) = \sum_{n=1}^{\infty} a_{2n} \frac{2n+1}{2n+2} |\psi|^{2n},
\end{align}
where $a_{n}$ are the coefficients in the \sas{continued expansion}{series} \sas{of}{ Eq.~(\ref{eqn:alpha}) defined by the Taylor expansion of Eq.~(\ref{eqn:LIA2})}. Thus, when $\tilde{\alpha}=0$ we have the LIA, which is an integrable theory on the geometric variables from the Frenet frame. As we will see, any amount of curvature correction to the integrable theory yields a non-Hamiltonian evolution. 

Our even expansions of $\alpha$ correct the cubic focusing nonlinear Schr\"odinger equation of LIA\sas{}{, to fourth order in $\kappa$,}  in the following way 
\begin{align}\label{eqn:GNVC}
i\psi_{t}+\psi_{ss} + \frac{1}{2} |\psi|^{2} \psi +
   a_2 \left(\left[|\psi|^{2} \psi\right]_{ss} + \frac{3}{4}|\psi|^{4}\psi\right)+a_4 \left(\left[|\psi|^{4} \psi\right]_{ss} + \frac{5}{6}|\psi|^{6}\psi\right) =0.
\end{align}
\sas{and indicates a straightforward pattern to arbitrary order, i.e., $a_{n}([|\psi|^{n} \psi]_{ss} + (n+1) |\psi|^{n+2}/(n+2))$.}{} While the \sas{power}{} nonlinearities \sas{}{due to powers of $|\psi|$} can adapt to the typical Hamiltonian structure of the integrable theory, the fully nonlinear \sas{terms}{derivative terms} cannot. Specifically, the question of whether the Hermitian inner-product on the Hilbert space of complex-valued square integrable functions on the real line induces a symplectic form such that one can identify a self-adjoint Hamiltonian whose variational derivative defines a Hamiltonian vector field consistent with Eq.~(\ref{eqn:GNVC}) has a negative answer.~[\onlinecite{Holmer2008GeometricEvolution}] Consequently, Noether's theorem is inapplicable and known symmetries need not generate conserved quantities.~[\onlinecite{user153764https://mathoverflow.net/users/110090/user153764InfinitesimalEvolution}] Our inability to formulate Eq.~(\ref{eqn:GNVC}) as an infinite-dimensional Hamiltonian flow is rooted to the fully nonlinear \sas{}{derivative} terms \sas{given by, $(|\psi|^{2n}\psi)_{ss}$}{}. Considering a \sas{linear}{smooth compactly supported} perturbation\sas{}{, $ \varepsilon \xi$,} of the \sas{lowest non-trivial order}{ fully nonlinear derivative term, $(|\psi|^{2}\psi)_{ss}$} we find \sas{}{the linearization}, 
\begin{align}\label{eqn:Ham1}
\left( |\psi + \varepsilon \xi |^{2} (\psi + \varepsilon \xi)\right)_{ss} = \left(|\psi|^{2} \psi\right)_{ss}+ \varepsilon \left(2|\psi|^{2} \xi + \bar{\xi}\psi^{2}\right)_{ss}+O(\varepsilon^{2}).
\end{align}
Assuming a smooth compactly supported perturbation and test function $u$ the functional given by the induced symplectic form yields~[\onlinecite{Buhler2006AMechanics}]
\begin{align}\label{eqn:Ham2}
\int  \bar{u}\left[2|\psi|^{2} \xi + \bar{\xi}\psi^{2}\right]_{ss}   ds = \int \bar{\xi}\left[2|\psi|^{2} u_{ss} + \bar{u}_{ss} \psi^{2}\right]   ds \neq \int \bar{\xi}\left[2|\psi|^{2} u + \bar{u} \psi^{2}\right]_{ss}   ds,
\end{align}
implying that a formal self-adjointness condition cannot be satisfied.~[\onlinecite{Olver2000ApplicationsEquations}] It can be verified that the linear derivative term and higher order power terms obey \sas{and}{a} Hamiltonian structure. Thus, our break from Hamiltonian structure is due to the full nonlinearity. Though our existing space-time symmetries do not yield the ``total energy'' and ``total momentum'' conservation typically associated with Schr\"odinger evolutions, this does not preclude the existence of conserved quantities nor additional non-obvious symmetries. However, application of the SYM symmetry software package~[\onlinecite{Dimas2006APDEs}] to the $a_{2}$ correction of Eq.~(\ref{eqn:GNVC}) found no additional continuous symmetries. Also, a Mathematica package that symbolically calculates conservation laws found no low-order conserved densities.~[\onlinecite{Poole2011SymbolicDimensions}]

\begin{figure}[h]
\includegraphics[width=\textwidth]{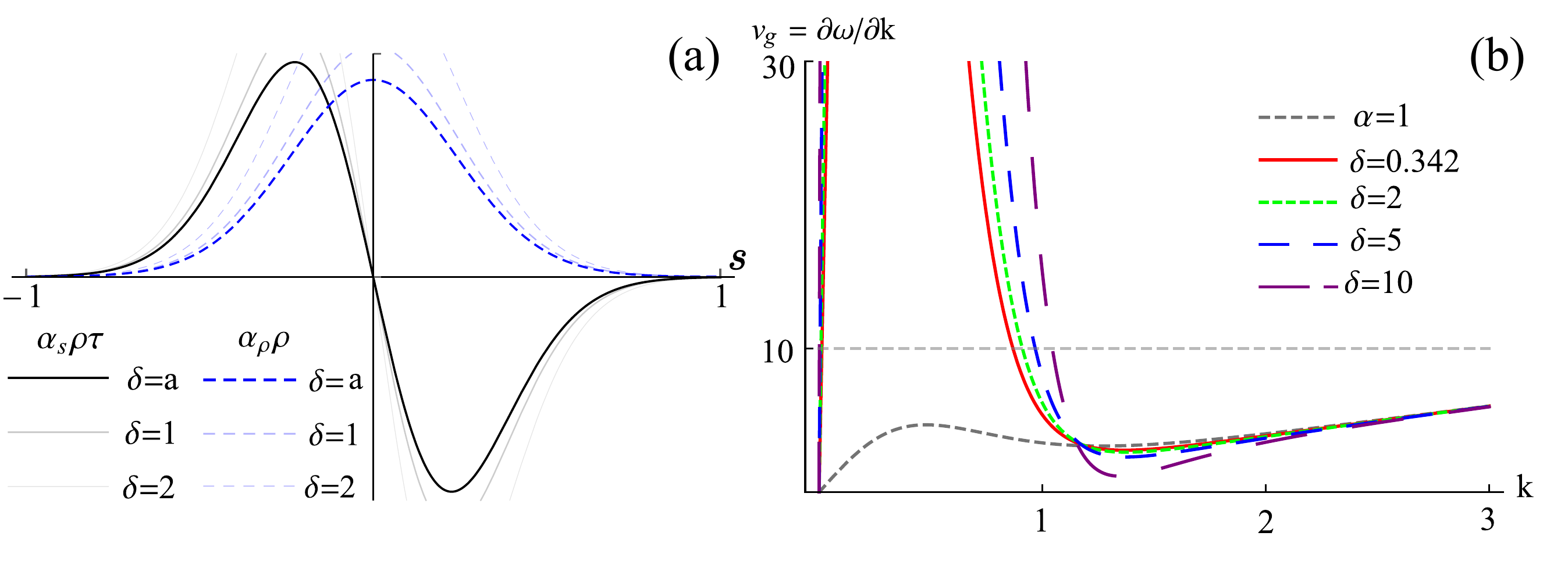}
\caption{Non-Hamiltonian gain/loss and dispersion of helical modes. Corrections to LIA increase/decrease line bending to right/left of a hyperbolic secant curvature profile with higher order corrections reinforcing this effect additively. Also, long wavelength modes propagate faster under correction causing a deformation of the initial curvature profile.  (a)  The integrand of Eq.~(\ref{eqn:totalBending1}), up through the first ten coefficients, for Hasimoto's vortex soliton (black.) The thinner and lighter black curves indicate how the integrand changes as we increase $\delta$. As we accumulate more vorticity with the Biot-Savart integral, the amplitude of the integrand increases while not distorting the basic shape. In addition, we plot the contribution due to the $\alpha_{\rho}\rho$ expansion of the integrand and see that the quantity is strictly positive. (b) The dispersion relation for helical modes of the initial state for the second correction. We see that the low wavenumber modes experience an enhanced dispersion, which grows as $\delta$ increases.}
\label{fig:dispersion}
\end{figure}

While this result speaks to the wholesale loss of Hamiltonian structure that appears as we move away from LIA, it tells us nothing about the wave motions of the vortex line. A useful perspective is given by decomposing the system into its real and imaginary components via Madelung's transformation to get, 
\begin{align}\label{eqn:Conservation}
\displaystyle \frac{\partial}{\partial t}\begin{bmatrix}
\rho \\ \tau
\end{bmatrix}
+ \frac{\partial}{\partial s}\begin{bmatrix}
2\alpha \rho \tau  \\ 
\displaystyle \left( \alpha \sqrt{\rho}\right)_{ss}/\sqrt{\rho} + f(\rho)\rho + \rho/2 - \alpha \tau^{2} 
\end{bmatrix} = 
\begin{bmatrix}
-2 \alpha_{s} \rho \tau  \\ 0 
\end{bmatrix},
\end{align}
where $\rho = \kappa^{2} = |\psi|^{2}$, which we call the {bending density}. If $\alpha=1$, then $a_{2}=0$ and we recover the standard hydrodynamic reformulation of cubic focusing NLS, which asserts that $\rho$ is conserved and $\tau$ obeys an Euler equation. Furthermore, curves of constant torsion define a Jacobian matrix whose spectrum reveals a single traveling wave solution which is the Hasimoto soliton or its generalization to elliptic representations of soliton trains.~[\onlinecite{Ludu2012NonlinearSurfaces}] The Jabcobian matrix in the general case, or for systems with non-constant torsion, are too complicated to analyze directly and we cannot make an assertion of hyperbolicity for the system. However, this reformulation does highlight the emergence of a source term in the bending density for non-constant $\alpha$. For compactly supported functions or those with suitable decay, we can integrate the first equation to find that \sas{}{for a segment of vortex line, parameterized by the arclength $s_0<s<s_1$,} the total bending \sas{across the vortex line} obeys 
\begin{align}\label{eqn:totalBending1}
\frac{d}{dt}\int_{s_0}^{s_1} \kappa^{2}ds = -2 \int_{s_0}^{s_1} \alpha_{s} \rho \tau ds = -2 \int_{s_0}^{s_1} (a_{2}+2a_{4}\rho + 3 a_{6} \rho^{2} +\cdots )\rho_{s}\rho \tau ds . 
\end{align}
Given that this non-conservation of the total norm can be traced back \sas{}{to} the fully nonlinear term in Eq.~(\ref{eqn:GNVC}), the previous loss of Hamiltonian structure is, perhaps, not surprising. It is interesting to note that if the coefficients $a_{n}$ are non-negative, then higher order corrections enter this formula additively and reinforce the gain/loss mechanism supplied by the first correction to LIA. \sas{}{Hasimoto originally considered a class of solitons defined by traveling curvature waves with constant torsion given by $\kappa = 2 \tau \mbox{sech}(\tau(s-ct))$ with $\tau = c/2$, which we refer to Hasimoto vortex solitons.} \sas{Considering  Hasimoto vortex soliton defined by $\kappa(s,t) = 2 \mbox{sech}(s-2t)$ with $\tau=1$,}{Setting $\tau=1$,} we calculate the \sas{}{parenthesis of the} last integrand\sas{}{, $\alpha_{\rho}$,} in Eq.~(\ref{eqn:totalBending1}) and find it to be non-negative, see Fig.~(\ref{fig:dispersion}a). This implies that, at least initially, the higher order contributions reinforce the gain/loss emergent in the first correction to LIA. Additionally, we plot the integrand, \sas{}{$\alpha_{s}\rho \tau$}, in Fig.~(\ref{fig:dispersion}a) and see that the vortex line should experience curvature gain \sas{to the right of}{ahead and loss behind} the soliton peak \sas{ and loss to the left}{which is propagating in the positive $s$ direction.} If we consider the system as a scalar conservation equation on $\rho$, for fixed $\tau$, then one can state the approximate characteristic speed as $c=[2\alpha \rho \tau]_{\rho} = 2\alpha_{\rho} \rho \tau + 2\alpha \tau$.~[\onlinecite{Toro2009NotionsEquations}] For $\alpha=1$, we have the LIA and a predicted speed of $c=2$ for the Hasimoto vortex soliton. Using our expansion for $\alpha$ provides the new approximate speed, $c= 2\tau( 1+ 2 a_{2} \rho + 3a_{4}\rho^{2} + 4a_{6} \rho^{3}+ \dots)$. As with Eq.~(\ref{eqn:totalBending1}), we see an additive influence of higher order corrections. In Sec.~(\ref{sec:simulations}) we consider simulations of Hasimoto's vortex soliton under a first correction to LIA. We find that the curvature peak, $\kappa_{\mbox{max}}$, has a strong linear relationship with the first correction, $\kappa_{\mbox{max}}(a_{2}) = 2.02810 -4.42020
a_{2}$, with a square residual of $0.9929$. Using this and a first correction of $c$ we find that the approximation quickly loses its accuracy with greater than a $3\%$ underestimation of the simulated peak speed for a $1\%$ correction strength. The implication is that the dynamics of gain/loss and torsion non-trivially affect the speed of the peak. Additionally, we see that if $\kappa \propto \mbox{sech}(s)$, then the approximate speed has a Gaussian-like profile which defines a non-convex/concave flux. \sas{}{That is, if the speed of a point on the traveling curvature wave is dependent on the value of curvature at that point, then the associated flux given by the first derivative of the speed with respect to $\rho$ shows that the speeds of points on the curvature distribution are neither strictly increasing nor decreasing functions of curvature. If, for example, the waves were strictly increasing with respect to curvature, then we would expect the soliton profile to undergo a wave steepening dynamic but this is not our case.} The simplest analog here is the Buckley-Leverett equation~[\onlinecite{Buckley1942MechanismSands}] which predicts \sas{}{a} shock front followed by a rarefaction wave. In our simulations, one can see a wake of helicity behind the propagating curvature profile. This structure is supported by the gain/loss mechanism which acts to evolve the soliton curvature profile to a step. However, this shock formation is tempered by other dynamics. In particular, there is a gross deformation of the curvature profile due to the nonlinear dispersion of helical/Fourier modes. 

A single mode helix, $\psi = \mathcal{A} e^{i(k s- \omega t)}$, which is a Fourier mode of the soliton state, initially obeys the nonlinear dispersion relation~[\onlinecite{Newton1987StabilityWaves}] associated with Eq.~(\ref{eqn:GNVC}),
 \begin{align}\label{eqn:Dispersion}
  \omega(k,\mathcal{A},\lambda) = k^{2}(1+ a_{2} \mathcal{A}^{2}+a_{4}\mathcal{A}^{4}+a_{6}\mathcal{A}^{6}) - \frac{\mathcal{A}^{2}}{2} - \frac{3}{4} a_{2} \mathcal{A}^{4} -\frac{5}{6}a_{4}\mathcal{A}^{6}  - \frac{7}{8} a_{6}\mathcal{A}^{8},
 \end{align}
where again we see the corrections alter the LIA dispersion relation additively. The Hasimoto vortex soliton given by $\psi = 2 \mbox{sech}(s) e^{is}$ contains  $95\%$ of its total Fourier \sas{mass}{energy} contained between wave numbers $k\in[-2,2]$.  Figure (\ref{fig:dispersion}b) plots the group velocity, where wave amplitude, $\mathcal{A}$, is related to wavenumber $k$ via Fourier transform, for the second correction to LIA. We find that these corrective terms seek to enhance the propagation speed of long wavelength modes, which will cause the initial curvature profile to distort. Thus, together with Eq.~(\ref{eqn:totalBending1}), the $a_{4}$ and $a_{6}$ corrections reinforce the gain/loss and dispersion mechanisms seen in the $a_{2}$ correction, a pattern which \sas{}{we checked} holds for the first 10 nontrivial corrections. As we reported in a previous work [\onlinecite{Strong2017Non-HamiltonianCondensates}] for a perturbative correction, the original peak can maintain localization even under the enhanced dispersion. Thus, when the kink is discernible, it is reasonable to consider it  a dissipative soliton.~[\onlinecite{Akhmediev2005DissipativeSolitons}]
\section{\label{sec:simulations} Simulating Binormal Vortex Motion}

Together, continuum mechanics, vector analysis,\sas{}{and} Helmholtz's and Kelvin's theorems from fluid mechanics assert that the motion of a vortex line is prescribed by the flow of the ambient field in which it is embedded. Past the lowest order approximation, the dynamics are sufficiently complicated \sas{and}{to} necessitate\sas{}{s} the use of numerical tools. The previous sections imply two distinct simulation procedures. The first is clear cut and relies on the approximation of solutions to initial-boundary value problems evolved according to an approximation to the vector evolution, Eq.~(\ref{eqn:vortex}). With the existence of efficient routines to evaluate incomplete elliptic integrals~[\onlinecite{Fukushima2011PreciseTransformations}] it appears possible to simulate the binormal evolution outright without approximation, however, such an investigation has never been attempted. Instead people work through the full BSI over an interpolated mesh or approximations via LIA at mesh points.  We call simulations stemming from Eq.~(\ref{eqn:binormal2}), {vector \sas{evolutions/}{}simulations} and consider first and second corrections to LIA given by Eq.~(\ref{eqn:BNF})\sas{}{.} 

On the other hand, the Hasimoto transform works by separating the parameterization of the vortex line from the evolution of its intrinsic geometric description given by the curvature and torsion variables. We will call a simulation of the vortex through the \sas{of the}{} geometric variables a {Hasimoto \sas{evolution/}{}simulation}. Naturally, this procedure introduces a post-processing step, which reconstructs the curve through the Frenet-Serret equations.~[\onlinecite{Bloch1997AGeometry}] Specifically, we must find the tangent vector by solving the following non-autonomous linear system of equations 
\begin{align}\label{eqn:FS}
\frac{d}{ds}\begin{bmatrix}{\textbf{T}} \\ {\textbf{N}}\\ {\textbf{B}} \end{bmatrix}  = \begin{bmatrix} 0 & \kappa & 0 \\ -\kappa & 0 & \tau \\ 0 & -\tau  & 0 \end{bmatrix} \begin{bmatrix}{\textbf{T}} \\ {\textbf{N}}\\ {\textbf{B}}. \end{bmatrix} 
\end{align}
From the tangent vector the curve's parameterization is found. The coefficient matrix in Eq.~(\ref{eqn:FS}) is skew-symmetric and thus an infinitesimal generator of the rotations mapping the Frenet frame from one point on the vortex line to the next. More importantly, these group elements of $SO(3)$ have spinor representations in the general linear group of two-by-two complex valued matrices under the isometric mapping,  $-2 x_{k} =\Tr({X}{\sigma}_{k}),\, \,  k=1,2,3,$ where ${X} = i(x_{1} \mathbf{\sigma}_{z}-x_{2}{\sigma}_{y}-x_{3}{\sigma}_{x})$ are defined through the standard Pauli spin matrices. In this representation, the Frenet-Serret equations are implicitly defined by the lower dimensional form,
\begin{align}\label{eqn:FSSpinor}
 \frac{d{U}}{ds} = {F}(s) {U}, \quad 2{F}(s) = \begin{bmatrix} 0 & i\psi/2 \\ i \bar{\psi}/2 & 0 \end{bmatrix},
\end{align}
where $\psi = \kappa \, \mbox{Exp}\left[ \int_{0}^{s} ds' \tau \right]$ and ${U}\in\mathbb{C}^{2 \times 2}$ such that $\bar{{U}}^{\textsc{t}}  {U}={I}$, which defines the spinor tangent vector ${E}_{1}=i{U}^{-1}{\sigma}_{z} {U}$.~[\onlinecite{Grinevich1997ClosedEquation},\onlinecite{Burns1991ModernFields}]  Interestingly, the spinor representation yields a Frenet-Serret coefficient matrix where the curvature and torsion are cast into the form of Hasimoto's wave function. Thus, the second process is to simulate the evolution of vortex line configurations through  Eq.~(\ref{eqn:GNVC}) and then recover the curve geometry by the application of quadrature to the traced out tangent vector created by the numerical approximation of Eq.~(\ref{eqn:FSSpinor}). It is also possible to numerically differentiate rectified phase data to recover the torsion, which can then be used in Eq.~(\ref{eqn:FS}).

We consider the $a_{2}$ correction to the evolution of a Hasimoto vortex soliton given by the initial state $\psi(s,0) = 2 \mbox{sech}(s) e^{is}$. To gain intuition over how a perturbation affects the solitonic evolution, we consider three cases\sas{}{,} $a_{2} \in \left\{0.11,0.19,0.23\right\}$. When $a_{2}=0.11$ a small number of low wavenumber curvature modes, $13.40$\% of the total Fourier energy, begin to propagate faster than those modes responsible for 95\% of the total initial bending. At $a_{2}=0.19$, $38.55$\% of the total Fourier energy is contained in the low wavenumber modes propagating faster than the remaining modes. The final value is chosen so that 47.73\% of the initial bending is propagating faster than the remaining modes. Density plots of these three cases are given in Fig.~(\ref{fig:DPHasimoto transform}). In each \sas{}{case}, we see an asymmetric evolution consistent with the gain/loss mechanism described in Sec.~{\ref{sec:Hasimoto transform}. We also\sas{,}{} see that \sas{as}{} dispersion of low wavenumber modes smears the distribution out\sas{}{.} \sas{and the peak is not completely eroded, which is due to the support it receives from the non-Hamiltonian gain mechanism.}{Additionally, the non-Hamiltonian gain mechanism keeps the peak from being completely eroded by dispersion.} \sas{Additionally, we see that f}{F}or smaller corrections, there is a breathing feature, which causes pockets of small curvature to form tightening the localization of the curvature peak. This feature, which was confirmed with an analysis of the power spectrum, is short-lived  under strong dispersion. In Fig.~(\ref{fig:vortexconfigA2}) we depict vortex lines produced from this correction to LIA at $t=10$. Figures (\ref{fig:vortexconfigA2}a)-(\ref{fig:vortexconfigA2}c) depict vortex lines corresponding to  the density plots in Fig.~(\ref{fig:DPHasimoto transform}). The remaining two are the result of higher order correction and we see that the Hasimoto vortex soliton decomposes itself into helical excitations of the vortex line. In light of the way corrections appear additively in speed, bending and dispersion calculations, it is reasonable that the evolution to a Kelvin wave cascade is more profound at higher order.  
\begin{figure}
 \includegraphics[width=\textwidth]{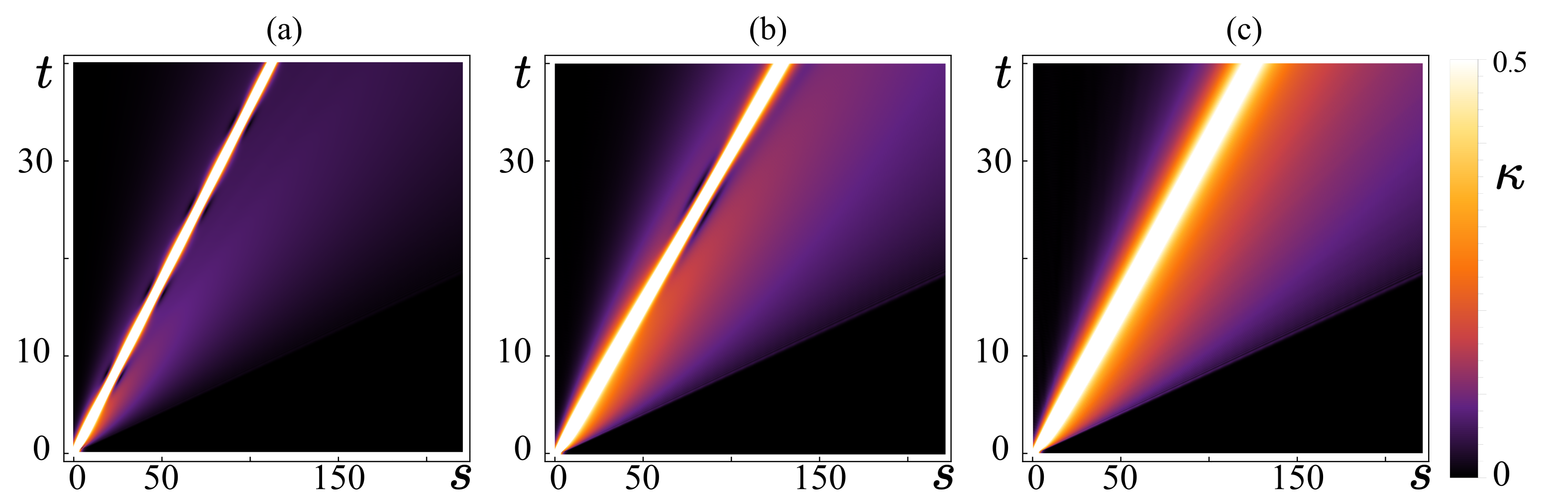}
\caption{Density plots of Hasimoto vortex soliton under first correction. The non-Hamiltonian binormal evolution of the bright soliton state maintains the kink feature despite dispersive deformations to the curvature distribution. (a) Evolution of the initial soliton state for $a_{2}=0.11$ produces a tightly confined peak with limited dispersion to the right. A slight breathing fluctuation is present. (b) The first correction is now $a_{2}=0.19$, and we see increased dispersion and a less frequent but more pronounced breathing dynamic. (c) Lastly, we have $a_{2}=0.23$ and see that under strong dispersion the breathing abates but a propagating peak, supported by the non-Hamiltonian gain mechanism, remains. }
\label{fig:DPHasimoto transform}
\end{figure}
\begin{figure}[h]
\includegraphics[width=\textwidth]{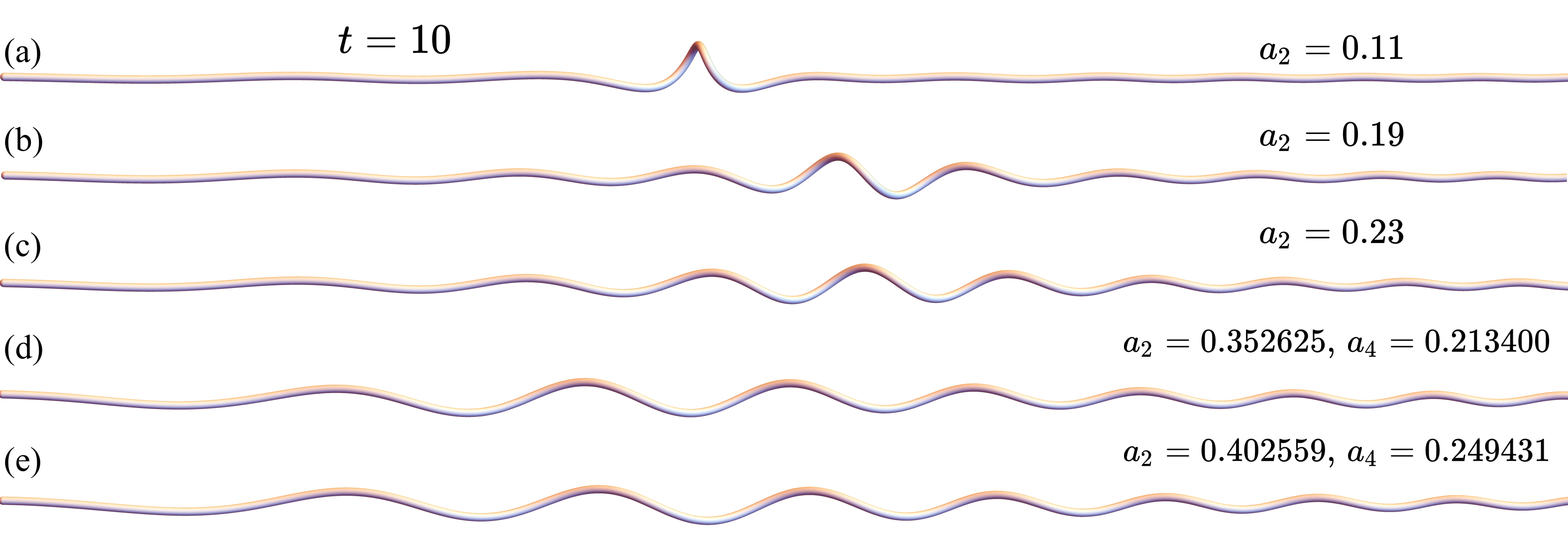}
\caption{Vortex configurations. Subplots (a)-(c) correspond to the evolutions in Fig.~(\ref{fig:DPHasimoto transform}) at $t=10$. We see that the increased dispersion corresponds to Kelvin wave generation, the largest in amplitude of which is an artifact of the original kink. (d) A Hasimoto soliton under the second correction where the coefficients, $a_{2} = 0.352625$ and  $a_{4}= 0.213401$, are chosen so that $\delta \to a$, which represents the smallest length of regularized vortex permitted under the Bustamante-Nazarenko cutoff. The quick decomposition of the soliton into a Kelvin wave cascade pronounced. (e) Hasimoto soliton evolved with $a_{2} = 0.402559$ and $a_{4}=0.249431$ which corresponds to $\delta =1$. We see a minimal change to the dynamics even though $\delta \approx 3 a$ indicating the corrections to LIA imply a rapid cascade dynamic.}
\label{fig:vortexconfigA2}
\end{figure}

Prior to Hayder's work of 2014~[\onlinecite{Salman2013BreathersVortices}], a comparison between Hasimoto, vector, and mean-field simulations had not been conducted. His work showed a qualitative agreement between the methods, except at points of reconnection which the mean-field model handled natively. In Fig.~(\ref{fig:HayderCurvatures}) we depict the $a_{2}$ correction to LIA, at various strengths, applied to the Akhmediev breather considered in~[\onlinecite{Salman2013BreathersVortices}]. We see that the correction not only increases the frequency of breathing, but the dispersion retards the relaxation to a non-peaked state. That the peaks are still maintained through several cycles is due to the non-Hamiltonian gain/loss mechanism.  The appearance of small-scale structures, caused by  wave interference across the periodic boundary, resulted in inefficient simulations for larger corrections. Using the spinor representation of the Frenet-Serret equations we were able to reconstruct the vortex line and found results consistent with the Hasimoto vortex solitons applied to each loop formed by the twisting and bending of the breathing dynamic. Specifically, the traveling curvature events, emerging from \sas{a breath}{one period of the breathing}, jettison helical excitations cascading Kelvin waves \sas{energy} away from the regions of highly localized curvature making it impossible to achieve a full exhalation, see from Fig.~(\ref{fig:HayderCurvatures}b). 
 \begin{figure}
\includegraphics[width=\textwidth]{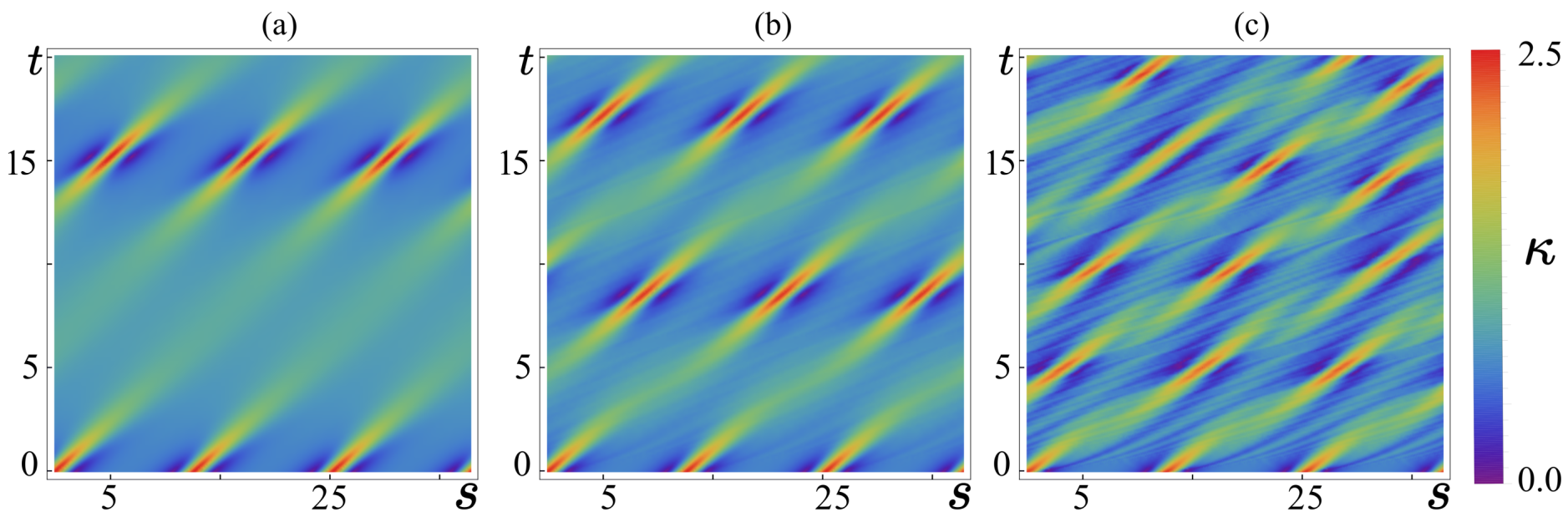}
\caption{Evolution of space-time periodic Akhmediev breather. (a) We plot the LIA evolution of the breather first considered in~[\onlinecite{Salman2013BreathersVortices}]. (b) The first correction to LIA, where $a_{2}=0.01$, increases the frequency of breathing. (c) Increasing the strength of correction to $a_{2} = 0.05$ sees two major changes. First, the frequency of breathing continues to increase with correction strength. Second, the dispersion tends to erode the relatively flat period \sas{in-between breaths}{occurring during one period of breathing}.}
\label{fig:HayderCurvatures}
\end{figure}

Lastly, we conducted simulations on vortex rings perturbed by Kelvin waves and vortex rings with initially localized \sas{bumps}{out of plane perturbation to a vortex ring}, similar to those seen after reconnection in classical hydrodynamics. Under LIA, \sas{rings with bumps}{perturbed rings} oscillated about the plane normal to their direction of motion as the \sas{bump}{perturbation} release\sas{d}{s} its bending into the ring in the form of \sas{wavy perturbations}{smaller amplitude traveling helical waves.} \sas{of the ring structure.}{} Initial testing indicates that \sas{a bump will}{the perturbation} create\sas{}{s} two curvature disturbances that are similar to a Hasimoto soliton traveling around the ring. However, these features were not true soltions and lost their shape as they traveled, even under LIA. A first correction to LIA increased the speed of propagation of the ring and helical decomposition of any kinks formed on the ring, see Fig.~(\ref{fig:ringDiagram}). On the other hand, the Kelvin rings which were tested appeared stable under LIA and corrected LIA. Specifically, while the speed of rotation and propagation was enhanced, the shape was relatively un-deformed when compared against LIA. While further testing is necessary, the possible decomposition of \sas{bumps}{perturbations} into Kelvin rings may provide stability to closed vortex structures propagating through mean-field simulated BEC.   
\begin{figure}[h]
\includegraphics[width=\textwidth]{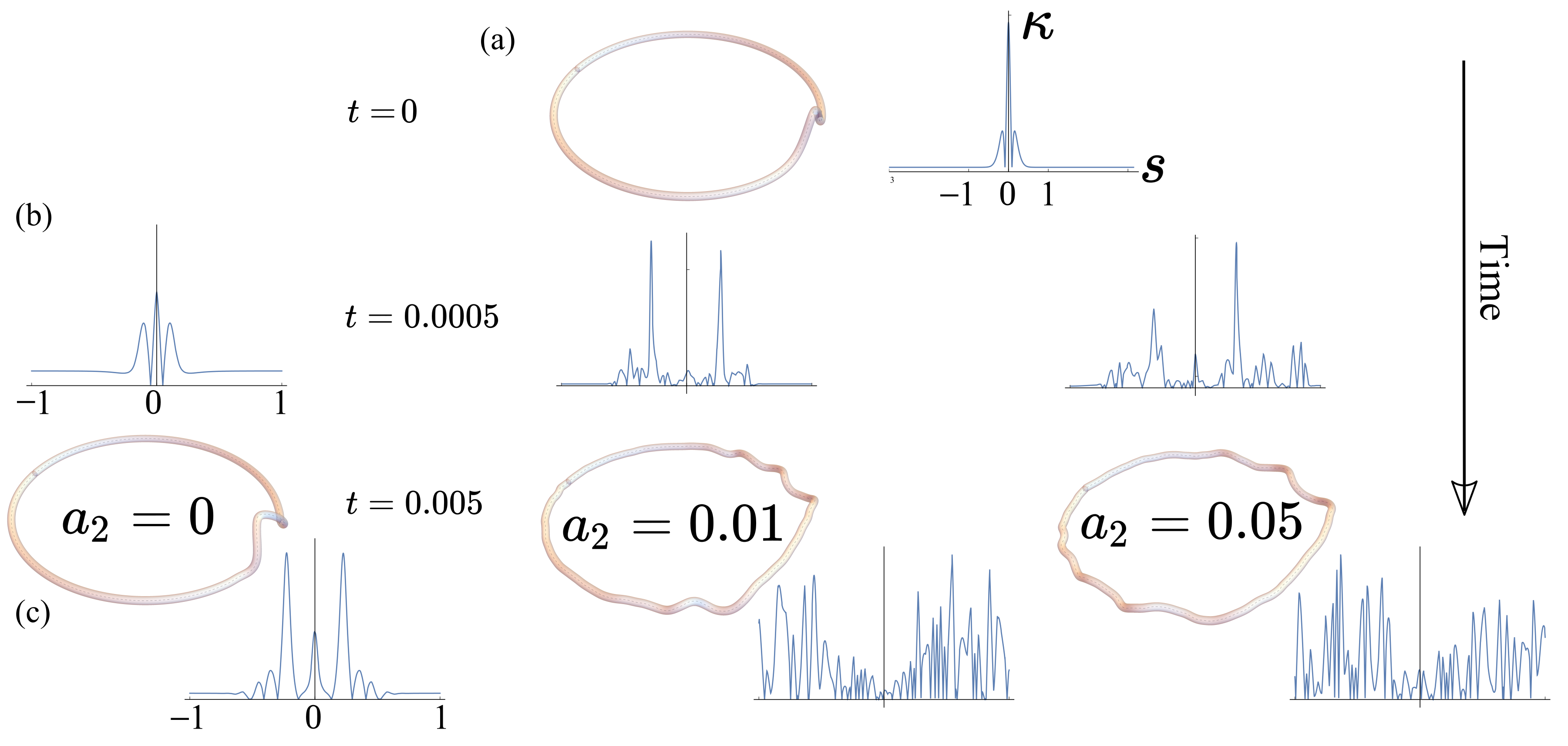}
\caption{Perturbed Rings. (a) A \sas{bump}{non-planar perturbation} on a vortex \sas{}{ring}, whose height is one-third of the ring radius, and its curvature profile. (b) Curvature profiles for the three evolutions, LIA where $a_{2}=0$, $a_{2}=0.01$, and $a_{2}=0.05$ at $t=0.0005$ shows that the \sas{bump}{perturbation} jettisons curved regions away from the initial deformation.   (c) Plots of the initial vortex state under LIA and corrections. We see that the corrections greatly increases the speed of the dynamic. For LIA we see the emergence of two curvature peaks propagating \sas{of}{away from} the initial peak. The corrections produce a similar state in shorter time. Fast moving waves vibrate the ring and cause noise in the curvature distributions. That said, in the vortex shows the existence of two counter propagating kink formations. All of these dynamics seek to distribute the bending along the vortex in a way similar to that which is seen in interacting bubble rings in classical hydrodynamic settings.}
\label{fig:ringDiagram}
\end{figure}

\section{\label{sec:conclusions} Discussion and Conclusions}

In this paper we present\sas{}{ed} an integrability breaking modification to the local induction approximation that maintains the \sas{Killing symmetry}{arclength preserving Hamiltonian structure} of the Eulerian flow induced by a vortex line while enhancing dispersion and introducing a non-Hamiltonian gain/loss mechanism affecting the geometric properties of the vortex medium.  This correction allows localized curvature distributions to decompose into Kelvin wave packets. In fact, we derive a hierarchy of non-Hamiltonian vortex cascade evolutions, which limit to an integro-differential equation resulting from the Hasimoto transformation of arbitrary curvature dependent binormal flows defined by the Biot-Savart representation of the velocity field induced by a vortex line. While we are unable to prove positivity in the coefficient structure of our expansion, which is an open and hard mathematical problem, the first ten non-trivial coefficients do not contradict this  conjecture and a Domb-Sykes analysis predicts a radius of curvature for our expansion which is consistent with the assumptions of our derivation. If the coefficients are positive to all orders, then the higher order terms in the asymptotic expansion of the local field additively reinforce the emergent non-Hamiltonian dynamics given by the first correction to the lowest order integrable theory. In other words, all curvature driven non-stretching Eulerian evolutions of vortex lines seek to disperse locally bent regions by the generation of helical waves. \sas{}{Additionally, in the setting of Kolmogorov-Arnold-Moser theory, the implication of the break in LIA integrability contrasted with the conservation law maintained by binormal flow is unclear since not every Euler evolution is an integrable one. Future connections to the differential geometry of fluid flows may shed light on this interesting and open question.}

The coefficient formulae depend on the characteristic length scales defined by arclength, vortex core size and local curvature and are ready for integration into established vortex filament methods.~[\onlinecite{Hanninen2014VortexTurbulence.}] These methods are used to simulate quantum fluids with a dense arrangement of quantized vortices and gain efficiency by approximation of the Biot-Savart integral with locally induced flows. They result in qualitatively similar dynamics for systems where vortex-vortex and self-interaction \sas{interaction}{} is weak.~[\onlinecite{Salman2013BreathersVortices}] When these dynamics dominate the flow, mean-field models that take into account \sas{the}{additional} physics of the vortex core must be included. That said, our analysis is appropriate for flows induced by lengths of vorticity nearing these scales and may represent as far a regulated Biot-Savart line integral can take the model into a core structure. With the emergence of experiments at both larger~[\onlinecite{Note1}] and small scale\sas{}{s}, focused on tangle behavior and primitive vortex interactions~[\onlinecite{Serafini2017VortexCondensates}], the importance of efficient multi-scale models for quantum turbulence is clear.~[\onlinecite{Hanninen2014VortexTurbulence.}]   

Preliminary analysis and simulations show that local but non-integrable induction models permit the excitation of Kelvin waves and indicate that a vortex line may attempt to find stability by storing bending in helical coils arrived at by a curvature cascade. It is expected that this energy transfer process couples to the fluid so that the bending can be relaxed through long wavelength acoustic fluctuations of the mass density. Incorporation of this effect into the geometric picture supports a connection between geometric hydrodynamics and the analysis of anomalous dissipation  conjectured by Onsager.~[\onlinecite{Eyink2006OnsagerTurbulence},\onlinecite{Onsager1931ReciprocalI.},\onlinecite{Onsager1931ReciprocalII.}] The current understanding is that classical Navier-Stokes solutions exhibiting anomalous kinetic energy dissipation in the inviscid limit correspond to weak Euler solutions, the singular fields defined by vortex lines being one such example.~[\onlinecite{Drivas2017AnEquations}] Furthermore, this mechanism is expected to be non-existent above a certain degree of solution regularity. The Biot-Savart perspective may be compatible with the regularity analysis associated with Onsager's conjecture and provide a geometric connection between anomalous dissipation and the relaxation of  ultracold quantum turbulence observable through the wave motion of the vortical substructure. 

A less theoretical application can be found in the recent high-resolution imaging of vortex ring breakdown where a self-similar decomposition of toroidal rings into vortex filaments occurs and is very much in the spirit of the predictions of Richardson and da Vinci. High-speed and high-resolution imagery~[\onlinecite{McKeown2017TheCollisions}] shows that interactions leading to tent formations~[\onlinecite{Kimura2018ScalingEvolution},\onlinecite{Kimura2018AEvolution}] and the flattening of tubes leads to filament generation at fine scales. In other words, the smoke we see as classical rings collide hides a discernible vortical skeleton comprised of bent filaments. At the finest scales, vortex lines should be the most appropriate model for the dynamics of the bent classical filaments and offer an opportunity to provide experimental corroboration of line models. In a similar thread, recent experiments seek to induce vortex line interactions in Bose-Einstein condensates whose vortical structure is not dense.~[\onlinecite{Serafini2017VortexCondensates}] This offers the clearest picture of vortex line dynamics, post tent formation and will certainly be an important apparatus in understanding primitive vortex dynamics. As our understanding of Kelvin wave generation on vortex lines and filaments is still in its early stages, such theoretical/experimental crossovers will be important for the continued development of future theories.

The authors would like to acknowledge Willy Hereman, Igor Khavkine, Randy LeVeque, Stephen Pankavich,  Barbara Prinari and David Sommer for useful discussions.  The authors acknowledge support from the US National Science Foundation under grant numbers PHY-1306638, PHY-1520915, OAC-1740130, and the US Air Force Office of Scientific Research grant number FA9550-14-1-0287. This work was performed in part at the Aspen Center for Physics, which is supported by the US National Science Foundation grant PHY-1607611. 


%

\end{document}